\documentclass[pra,superscriptaddress,longbibliography,twocolumn]{revtex4-2}
\usepackage[utf8]{inputenc}
\usepackage[english]{babel}
\usepackage{amsmath}
\usepackage{amsfonts}
\usepackage{amssymb}
\usepackage{graphicx}
\usepackage{amsthm}
\usepackage{color}
\usepackage{hyperref}
\usepackage[caption=false]{subfig}
\usepackage{graphicx,float}
\usepackage{dcolumn}
\usepackage{bm}
\usepackage{mathrsfs}
\usepackage{txfonts}
\usepackage{CJK}
\usepackage[amssymb]{SIunits}
\usepackage{epsfig}
\usepackage{epstopdf}
\usepackage{lipsum}
\usepackage{placeins}
\usepackage{physics}
\usepackage{mathtools}
\usepackage{blindtext}
\newenvironment{SChinese}{%
\CJKfamily{gbsn}%
\CJKtilde
\CJKnospace}{}
\allowdisplaybreaks[2]

\begin{document}

\begin{CJK}{UTF8}{}
\begin{SChinese}

\title{Nonlinear dissipation induced photon blockade}%

\author{Xin Su (苏欣)}
 \affiliation{College of Engineering and Applied Sciences, and National Laboratory of Solid State Microstructures, Nanjing University, Nanjing 210093, China}
 \affiliation{School of Electronic Science and Engineering, Nanjing University, Nanjing 210093, China}

\author{Jiang-Shan Tang (唐江山)}%
\email{js.tang@foxmail.com}
  \affiliation{College of Engineering and Applied Sciences, and National Laboratory of Solid State Microstructures, Nanjing University, Nanjing 210093, China}
  \affiliation{School of Physics, Nanjing University, Nanjing 210023, China}

\author{Keyu Xia (夏可宇)}  %
 \email{keyu.xia@nju.edu.cn}
    \affiliation{College of Engineering and Applied Sciences, and National Laboratory of Solid State Microstructures, Nanjing University, Nanjing 210093, China}

\date{\today}

\begin{abstract}
We theoretically propose a scheme for photon blockade in a cavity quantum electrodynamical system consisting of an N-type atomic medium interacting with a single-mode Fabry-P\'{e}rot cavity. In contrast to inefficient nonlinear-dispersion-induced photon blockade suppressed by a large detuning, the photon blockade in our scheme is induced by a large nonlinear dissipation of the cavity created by the N-type atomic system. A deep photon blockade is manifested with a vanishing equal-time second-order correlation function within the cavity linewidth. This work provides an efficient photon blockade because it work in the near-resonance case.
\end{abstract}

\maketitle

\end{SChinese}
\end{CJK}

\section{\label{sec:level1}Introduction}
Manipulation of single photons has been one of the key tasks of quantum information science and technology. It can be linear operation on photons, such as the storage and state manipulation of photons \cite{QuantumMemory,Dark-StatePolariton,ObservationCoherentStorage,StorageInAtomicVapor,QuantumStateTransfer,EITSingle-photon,AtomicMemoryCorrelatedPhotons}, or nonlinear operations, such as the generation of single photons via photon blockade (PB) \cite{PBWithOneTrappedAtom,PhotonTurnstile,QuantumDotPhotonTurnstile,On-DemandHeraldedSingle-PhotonSources} and quantum logic gates \cite{MeasurementConditionalQPG,QED-basedQPG,photon-photonGate, CoherentOperationTunableQPG,QuantumGateFlyingPhotonTrappedAtom,QuantumLogicStark-shiftedRamanTransition}. The nonlinear control of photons requires strong photon-photon interaction, or conventionally a giant dispersive Kerr nonlinearity \cite{GiantKerrNonlinearityEIT,StronglyInteractingPhotons,LargeKerrNonlinearitySingleAtom,EITReview}, in which the phase of a signal field is modified by an amount proportional to the photon number in another field or its own photon number, with both of the two fields containing few photons. In addition to being utilized to prepare single photon sources or realize quantum logic gates, the giant optical Kerr nonlinearity also enable a great number of useful applications such as nondestructive measurement of photons \cite{QNDMeasurementViaKerrEffect,QNDMeasurementsInOptics,PhotonNumberSelectiveGroupDelay,NondestructiveDetectionOfPhoton} and single-photon switches and transistors \cite{All-OpticalSwitchGatedByOnePhoton,UltrafastSwitchingBySinglePhoton,Single-PhotonTransistorSurfacePlasmons,All-OpticalSwitchWithinHollowFiber,All-opticalRoutingByOne-atomSwitch}.

The dispersive Kerr nonlinearity of conventional materials is typically negligible at the single-photon level \cite{QuantumNonlinearOptics-PhotonByPhoton}. To tackle this problem, various quantum nonlinear systems such as the atom-cavity systems \cite{PBWithOneTrappedAtom,StronglyInteractingPhotons}, Rydberg atomic ensembles \cite{RydbergEnsemble,RydbergEnsemble2} and single quantum dots strongly coupling to a photonic crystal cavity \cite{QuantumDotCoupledToPC,ControlledPhaseShift-QuantumDotCoupledToPC}, have been proposed to generate strong photon-photon interactions. Among these systems, a cavity quantum electrodynamical system embedded with a N-type atomic system have been theoretically proposed \cite{GiantKerrNonlinearityEIT,StronglyInteractingPhotons,LargeKerrNonlinearitySingleAtom,EITReview,NonlinearOptics-LowLightLevel} and experimentally demonstrated \cite{ObservationLargeKerrNonlinearity,StrongCouplingPhotons,EIT-based_XPM_attojoule_levels,ReversibleSelfKerrNonlinearitySwitchingField} to induce a giant Kerr nonlinearity with vanishing one-photon absorption. This absorptive-free giant Kerr nonlinearity promises significant practical applications in photon blockade \cite{StronglyInteractingPhotons,PPICavityEIT,PhotonStatisticsEIT,StrongCouplingPhotons} and photonic chirality \cite{Cavity-FreeIsolatorCirculator,ReversibleSingle-photonIsolation}. Nevertheless, it is suppressed by a large detuning in the nonresonance case, in order to dominate the two-photon absorption. As a result, the efficiency of the PB induced by this dispersive nonlinearity is limited. Yet, in comparison with the dispersive Kerr nonlinearity, a larger two-photon absorption can be obtained with the N-type atomic system in the resonant case, giving rise to a stronger PB effect.

In this paper, we show a strong PB by virtual of the two-photon absorption of an optical cavity induced by the N-type atomic ensemble. In the conventional PB caused by a dispersive Kerr nonlinearity in the N-type system, the nonlinearity is suppressed by a large detuning. In contrast, the nonlinear dissipation in our scheme is obtained in the resonance case and thus greatly enhanced. Our scheme paves the way for using nonlinear dissipation as a new mechanism for the PB, which can be used to efficiently extract single photons from a weak coherent state of light.

Our paper is organized as follows. In Sec. \ref{sec 1: system and model}, we describe our system and the model, and use the perturbation method to analyse the effect of the N-type system on the Fabry-P\'{e}rot cavity: one-photon absorption, dispersion and two-photon nonlinearity. In Sec. \ref{sec 2: results}, we show the results for nonlinear dissipation induced PB and the transmission. In Sec. \ref{sec 3: implementation}, we present the experimental implementation for our proposal. In the end, we conclude our work in Sec. \ref{sec 4: conclusion}.

\section{System and model}\label{sec 1: system and model}

\begin{figure}
\centering
\includegraphics[width=1\linewidth]{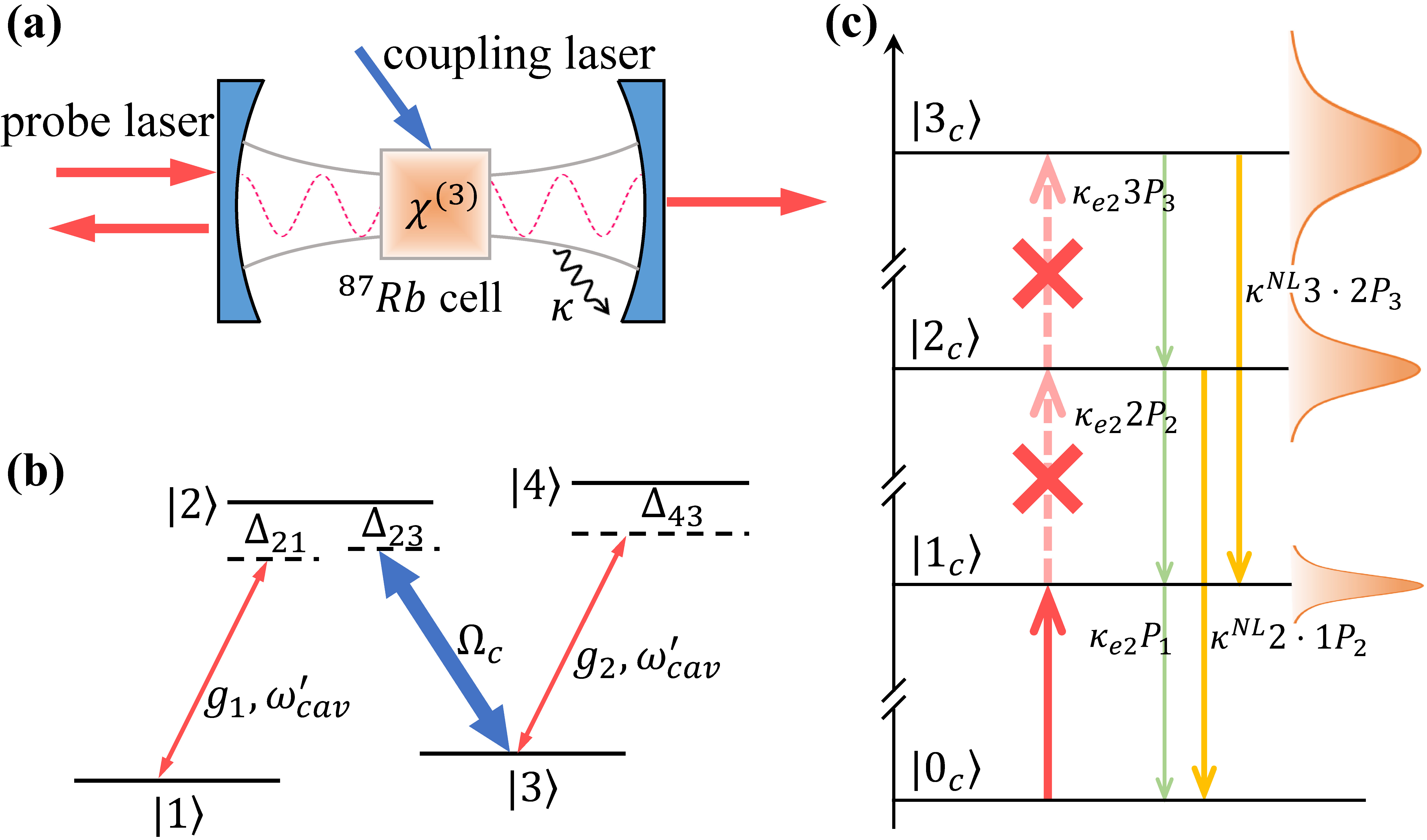}
\caption{(a) The schematic diagram of our system. A cell with an ensemble of $\prescript{87}{}{\mathrm{Rb}}$ atoms is placed in the cavity. (b) The related level structure of $\prescript{87}{}{\mathrm{Rb}}$ atoms. (c) All `intrinsic' dissipation channels besides the input coupling channel and nonlinear `intrinsic' dissipation rates for different Fock states.}
\label{fig:Fig1}
\end{figure}

Here, we explain our idea to achieve anti-bunched photons via the two-photon absorption in a single-mode two-sided cavity. It is assumed that there is no intrinsic one-photon absorption in the cavity. The cavity mode has a linear decay rate $\kappa_{e1}$ ($\kappa_{e2}$) for its coupling to the outside field at the input (second) cavity mirror. The change rates of Fock-state probabilities $P_n$ in this system is given by
\begin{equation}
\begin{split}
\dot{P}_{n}(t)=&\kappa^\mathrm{L}\left[(n+1) p_{n+1}-n p_n\right]+\\
&\kappa^\mathrm{NL}\left[(n+2)(n+1) p_{n+2}-n(n-1) p_n\right],
\end{split}
\end{equation}
where $\kappa^\mathrm{L}=\kappa_{e1}+\kappa_{e2}$ ($\kappa^\mathrm{NL}$) is the total linear decay rate (two-photon absorption rate). We summarize the two-photon absorption losses in the cavity as well as the transmission losses at the second mirror as all `intrinsic' dissipation channels behind the input mirror. Our idea is schematically shown in Fig. \ref{fig:Fig1}(c). For a two-photon absorption process dominating the linear decay at the transmission port, the intracavity multiple photons are absorbed by two-photon annihilation cascade (yellow arrows in Fig.~\ref{fig:Fig1}(c)) and thus dissipate to vacuum or one-photon states before they decay at the transmission mirror as a mixture of single-photon and multi-photon states (green arrows in Fig.~\ref{fig:Fig1}(c)). Whereas single photons only experience the decay of the transmission channel behind the input port. In this way, the multi-photon states can not exist in the cavity and thus are blocked, while single photons are transmitted through the cavity, which ultimately induce the PB.

Our system for PB is schematically shown in Fig.~\ref{fig:Fig1}(a). It consists of a Fabry-P{\'e}rot (FP) cavity and an ensemble of $\prescript{87}{}{\mathrm{Rb}}$ atoms. The related level structure of the $\prescript{87}{}{\mathrm{Rb}}$ atoms is depicted in Fig.~\ref{fig:Fig1}(b). The atom in the N-type configuration has two ground states denoted by $|1\rangle$ and $|3\rangle$, and two excited states denoted by $|2\rangle$ and $|4\rangle$. A strong coupling laser with frequency $\omega_c$ is applied to the $|3\rangle \leftrightarrow |2\rangle$ transition. The $|1\rangle \leftrightarrow |2\rangle$ and $|3\rangle \leftrightarrow |4\rangle$ transitions are coupled to the cavity mode simultaneously. A weak probe laser with frequency $\omega_p$ drives the cavity via the input mirror with rate $\kappa_{e1}$. The cavity field escape from the output mirror with rate $\kappa_{e2}$. We assume that the intrinsic decay rate of the cavity is $\kappa_i$. It is negligible in our system. For simplicity, we set $\kappa_{e1}=\kappa_{e2}$. Thus, the cavity field decays at a total rate of $\kappa=\kappa_{e1}+\kappa_{e2}+\kappa_i$. This decay rate $\kappa$ is referred to as the linear dissipation rate $\kappa^\mathrm{L}$ aforementioned.

The Hamiltonian of the system in the rotating frame takes the form
\begin{equation}\label{eq1:Rotated Hamiltonian}
\begin{split}
\hat{H}&=-\Delta\hat{a}^\dag\hat{a}+\sum_{j=1}^{N}\left[\Delta_{21}\hat{\sigma}_{22}^{j}+\left(\Delta_{21}-\Delta_{23}\right)\hat{\sigma}_{33}^{j}\right.\\
&\quad +\left(\Delta_{21}-\Delta_{23}+\Delta_{43}\right)\hat{\sigma}_{44}^{j}+ig_1\left(\hat{a}^\dag\hat{\sigma}_{12}^{j}-\hat{\sigma}_{21}^{j}\hat{a}\right)\\
&\quad\left.+ig_2\left(\hat{a}^\dag\hat{\sigma}_{34}^{j}-\hat{\sigma}_{43}^{j}\hat{a}\right)+i\left(\Omega_c^\ast\hat{\sigma}_{32}^{j}-\Omega_c\hat{\sigma}_{23}^{j}\right)\right]\\
&\quad+i\sqrt{\kappa_{e,1}}\varepsilon_p\left(\hat{a}^\dag-\hat{a}\right).
\end{split}
\end{equation}
The first term in Eq. \eqref{eq1:Rotated Hamiltonian} is the free Hamiltonian of the intracavity field with annihilation and creation operators $\hat{a}$ and $\hat{a}^\dag$, where $\Delta=\omega_p-\omega_\text{cav}$ is the detuning between the probe field and the bare cavity resonance. The first three terms in the brackets represent the internal energy of the $j$th atom with $\hat{\sigma}^{j}_{mn}\equiv|m_j\rangle\langle n_j|$ ($m,n=1,2,3,4$) being the population operators (for $m=n$) or the atomic raising and lowering operators (for $m\neq n$) of the $j$th atom, in which $\Delta_{21}=\omega_{21}-\omega_\mathrm{cav}-\Delta$ and $\Delta_{43}=\omega_{43}-\omega_\mathrm{cav}-\Delta$ are the detunings between the corresponding transitions and the probe field, and $\Delta_{23}=\omega_{23}-\omega_c$ is the detuning between the $|3\rangle\leftrightarrow|2\rangle$ transition and the coupling laser. The fourth (fifth) term in the brackets describes the coupling between the cavity mode and the $|1\rangle\leftrightarrow|2\rangle$ ($|3\rangle\leftrightarrow|4\rangle$) transition with the coupling rate $g_1$ ($g_2$). The last term in the brackets expresses the interaction between the coupling laser and the $|3\rangle\leftrightarrow|2\rangle$ transition, in which $\Omega_c=\mu_{23}E_c/2\hbar$ is the half Rabi frequency of the coupling laser. The last term describes the coupling between the monochromatic continuous-wave probe field and the cavity mode, the amplitude of the probe field being $\varepsilon_p=\sqrt{P/\hbar\omega_p}$ where $P$ is the input power.

The decay and dephasing of the system can be described by the Lindblad operator
\begin{equation}\label{eq1:Lindblad operator}
\begin{split}
\hat{\mathcal{L}}\hat{o}&=\frac{\kappa}{2}\left[2\hat{a}^\dag\hat{o}\hat{a}-\hat{a}^\dag\hat{a}\hat{o}-\hat{o}\hat{a}^\dag\hat{a}\right]+\\
&\quad\frac{\gamma_{nm}}{2}\sum_{j=1}^{N}\left[2\hat{\sigma}^j_{nm}\hat{o}\hat{\sigma}^j_{mn}-\hat{\sigma}^j_{nn}\hat{o}-\hat{o}\hat{\sigma}^j_{nn}\right],
\end{split}
\end{equation}
where $\gamma_{nm}=\left\{\Gamma_{41},\Gamma_{43},\Gamma_{42},\Gamma_{21},\Gamma_{23},\Gamma_{31}\right\}$ with $\Gamma_{nm}$ denoting the spontaneous decay or dephasing rates from the state $|n_j\rangle$ to the state $|m_j\rangle$, and $\hat{\sigma}_{mn}=\left\{\hat{\sigma}_{14},\hat{\sigma}_{34},\hat{\sigma}_{24},\hat{\sigma}_{12},\hat{\sigma}_{32},\hat{\sigma}_{13}\right\}$.

Incorporating the Lindblad operator into the Heisenberg's equations derived from Eq. \eqref{eq1:Rotated Hamiltonian}, the evolution of the system is given by quantum Langevin's equations:
\begin{subequations}\label{eq1:EOM}
\begin{align}
\label{eq1:EOM_cavity}\dot{\hat{a}}&=\left(i\Delta-\frac{\kappa}{2}\right)\hat{a}+g_1\sum_{j=1}^{N}\hat{\sigma}_{12}^{j}+g_2\sum_{j=1}^{N}\hat{\sigma}_{34}^{j}+\sqrt{\kappa_{e,1}}\varepsilon_p,\\
\label{eq1:EOM_1}\dot{\hat{\sigma}}_{11}^j&=\Gamma_{31}\hat{\sigma}_{33}^j+\Gamma_{21}\hat{\sigma}_{22}^j+\Gamma_{41}\hat{\sigma}_{44}^j+g_1\left(\hat{a}^\dag\hat{\sigma}_{12}^j+\hat{\sigma}_{21}^j\hat{a}\right),\\
\dot{\hat{\sigma}}_{22}^j&=-(\Gamma_{21}+\Gamma_{23})\hat{\sigma}_{22}^j+\Gamma_{42}\hat{\sigma}_{44}^j-\left(\Omega_c^\ast\hat{\sigma}_{32}^j+\Omega_c\hat{\sigma}_{23}^j\right)-\notag\\
&\quad g_1\left(\hat{a}^\dag\hat{\sigma}_{12}^j+\hat{\sigma}_{21}^j\hat{a}\right),\\
\dot{\hat{\sigma}}_{33}^j&=-\Gamma_{31}\hat{\sigma}_{33}^j+\Gamma_{23}\hat{\sigma}_{22}^j+\Gamma_{43}\hat{\sigma}_{44}^j+\left(\Omega_c^\ast\hat{\sigma}_{32}^j+\Omega_c\hat{\sigma}_{23}^j\right)\notag\\
&\quad+g_2\left(\hat{a}^\dag\hat{\sigma}_{34}^j+\hat{\sigma}_{43}^j\hat{a}\right),\\
\label{eq1:EOM_4}\dot{\hat{\sigma}}_{44}^j&=-(\Gamma_{42}+\Gamma_{43}+\Gamma_{41})\hat{\sigma}_{44}^{j}-g_2\left(\hat{a}^\dag\hat{\sigma}_{34}^j+\hat{\sigma}_{43}^j\hat{a}\right),\\
\dot{\hat{\sigma}}_{23}^{j}&=-\tilde{\gamma}_{23}\hat{\sigma}_{23}^j-\Omega_c^\ast\left(\hat{\sigma}_{33}^j-\hat{\sigma}_{22}^j\right)-g_1\hat{a}^\dag\hat{\sigma}_{13}^j+g_2\hat{a}^\dag\hat{\sigma}_{24}^j,\\
\label{eq1:EOM_6}\dot{\hat{\sigma}}_{14}^{j}&=-\tilde{\gamma}_{14}\hat{\sigma}_{14}^{j}+g_1\hat{\sigma}_{24}^{j}\hat{a}-g_2\hat{\sigma}_{13}^{j}\hat{a},\\
\label{eq1:EOM_7}\dot{\hat{\sigma}}_{12}^{j}&=-\tilde{\gamma}_{12}\hat{\sigma}_{12}^{j}-\Omega_c\hat{\sigma}_{13}^{j}-g_1\left(\hat{\sigma}_{11}^{j}-\hat{\sigma}_{22}^{j}\right)\hat{a},\\
\dot{\hat{\sigma}}_{13}^{j}&=-\tilde{\gamma}_{13}\hat{\sigma}_{13}^{j}+\Omega_c^\ast\hat{\sigma}_{12}^{j}+g_1\hat{\sigma}_{23}^{j}\hat{a}+g_2\hat{a}^\dag\hat{\sigma}_{14}^{j},\\
\dot{\hat{\sigma}}_{24}^{j}&=-\tilde{\gamma}_{24}\hat{\sigma}_{24}^j-\Omega_c^\ast\hat{\sigma}_{34}^j-g_1\hat{a}^\dag\hat{\sigma}_{14}^j-g_2\hat{\sigma}_{23}^j\hat{a},\\
\label{eq1:EOM_10}\dot{\hat{\sigma}}_{34}^{j}&=-\tilde{\gamma}_{34}\hat{\sigma}_{34}^j+\Omega_c\hat{\sigma}_{24}^j-g_2\left(\hat{\sigma}_{33}^j-\hat{\sigma}_{44}^j\right)\hat{a},
\end{align}
\end{subequations}
where $\tilde{\gamma}_{12}=i\Delta_{21}+\gamma_{12}$, $\tilde{\gamma}_{13}=i(\Delta_{21}-\Delta_{23})+\gamma_{13}$, $\tilde{\gamma}_{34}=i\Delta_{43}+\gamma_{34}$, $\tilde{\gamma}_{24}=i(\Delta_{43}-\Delta_{23})+\gamma_{24}$, $\tilde{\gamma}_{23}=-i\Delta_{23}+\gamma_{32}$, and $\tilde{\gamma}_{14}=i(\Delta_{21}-\Delta_{23}+\Delta_{43})+\gamma_{14}$. We define $\gamma_{mn}=(\Gamma_n+\Gamma_m)/2$ with $\Gamma_n$ ($\Gamma_m$) being the total decay rate of population out of level $|n\rangle$ ($|m\rangle$).

The atomic degrees of freedom can be adiabatically eliminated \cite{AdiabaticElimination}, under the assumption that the spontaneous decay rates of the atoms are much larger than the atom-field coupling rates, i.e., $\Gamma_{nm}\gg g_1,\,g_2$, so that the atomic coherence and population operators $\hat{\sigma}_{mn}^j$ evolve much faster than $\hat{a}$ and reach their steady state much earlier than $\hat{a}$. Solving the steady-state solutions of $\hat{\sigma}_{12}^{j}$ and $\hat{\sigma}_{34}^{j}$ allow us to express the effective Hamiltonian and the additional decay process induced by atomic effects in terms of the mode operators. Under the condition that $g_1\langle\hat{a}\rangle,\,g_2\langle\hat{a}\rangle\ll\Omega_c$, the atomic operators can be perturbatively expanded \cite{PerturbationMethod} as
\begin{equation}
\hat{\sigma}_{mn}=\hat{\sigma}_{mn}^{(0)}+\hat{\sigma}_{mn}^{(1)}+\hat{\sigma}_{mn}^{(2)}+\hat{\sigma}_{mn}^{(3)}+\cdots,
\end{equation}
where we neglect the superscript ``$j$'' denoting the $j$-th atom for convenience. Since the cavity mode couples weakly to each individual atom, each atom can be assumed to be populated at the ground state to zeroth order, i.e., $\hat{\sigma}_{11}^{(0)}=1$, $\hat{\sigma}_{nn}^{(0)}=0\,(n=2,3,4)$, $\hat{\sigma}_{mn}^{(0)}=0\,(m\neq  n)$. Under this assumption, we iteratively determine the remaining components of higher orders in the expansion \cite{PerturbationMethod}. Substituting the zero order population and coherence into Eqs. \eqref{eq1:EOM_7}--\eqref{eq1:EOM_10} and Eqs. \eqref{eq1:EOM_1}--\eqref{eq1:EOM_6} respectively, the first-order perturbation of the coherence and population operators can be obtained to be
\begin{subequations}
\begin{align}
\label{eq1:first-order_sigma12}\hat{\sigma}_{12}^{(1)}&=\frac{-g_1\hat{a}}{\tilde{\gamma}_{12}+\frac{|\Omega_c|^2}{\tilde{\gamma}_{13}}},\\
\label{eq1:first-order_sigma34_sigma24}\hat{\sigma}_{34}^{(1)}&=\hat{\sigma}_{24}^{(1)}=0,\\
\hat{\sigma}_{nn}^{(1)}&=\hat{\sigma}_{23}^{(1)}=\hat{\sigma}_{14}^{(1)}=0\,(n=1,2,3,4).
\end{align}
\end{subequations}
In a closed atomic system, the total population is conserved, i.e., $\hat{\sigma}_{11}+\hat{\sigma}_{22}+\hat{\sigma}_{33}+\hat{\sigma}_{44}=1$. Then, it can be deduced directly that
\begin{equation}\label{eq1:conserved population}
\hat{\sigma}_{11}^{(2)}+\hat{\sigma}_{22}^{(2)}+\hat{\sigma}_{33}^{(2)}+\hat{\sigma}_{44}^{(2)}=0.
\end{equation}
Substituting Eqs. \eqref{eq1:first-order_sigma12}--\eqref{eq1:first-order_sigma34_sigma24} into Eqs. \eqref{eq1:EOM_1}--\eqref{eq1:EOM_4}, and combining with Eq. \eqref{eq1:conserved population} and the assumption of $\hat{\sigma}_{23}^{(2)}=\hat{\sigma}_{14}^{(2)}=0$, the second order correction can be obtained as
\begin{subequations}
\begin{align}
\label{eq1:second-order_sigma22}\hat{\sigma}_{22}^{(2)}&=\frac{g_1^2}{(\Gamma_{21}+\Gamma_{23})}\frac{2\mathrm{Re}[F_1]}{|F_1|^2}a^\dag a,\\
\hat{\sigma}_{33}^{(2)}&=g_1^2\frac{\Gamma_{23}}{\Gamma_{31}(\Gamma_{21}+\Gamma_{23})}\frac{2\mathrm{Re}[F_1]}{|F_1|^2}a^\dag a,\\
\hat{\sigma}_{11}^{(2)}&=-g_1^2\frac{\Gamma_{23}+\Gamma_{31}}{\Gamma_{31}(\Gamma_{21}+\Gamma_{23})}\frac{2\mathrm{Re}[F_1]}{|F_1|^2}a^\dag a,\\
\label{eq1:second-order_sigma44}\hat{\sigma}_{44}^{(2)}&=0,\\
\label{eq1:second-order_coherence}\hat{\sigma}_{mn}^{(2)}&=0\,(m\neq n),
\end{align}
\end{subequations}
where $F_1=\tilde{\gamma}_{12}+|\Omega_c|^2/\tilde{\gamma}_{13}$. Similarly, substituting the second-order perturbation of the operators Eqs. \eqref{eq1:second-order_sigma22}--\eqref{eq1:second-order_coherence} into Eqs. \eqref{eq1:EOM_7}--\eqref{eq1:EOM_10}, we obtain $\hat{\sigma}_{12}$ and $\hat{\sigma}_{34}$ to third order
\begin{subequations}
\begin{align}
\label{eq1:third-order_sigma12}\hat{\sigma}_{12}^{(3)}&=g_1^3\frac{\Gamma_{23}+2\Gamma_{31}}{\Gamma_{31}(\Gamma_{21}+\Gamma_{23})}\frac{2\mathrm{Re}[F_1]}{F_1|F_1|^2}\hat{a}^\dag\hat{a}^2,\\
\label{eq1:third-order_sigma34}\hat{\sigma}_{34}^{(3)}&=-g_2g_1^2\frac{\Gamma_{23}}{\Gamma_{31}(\Gamma_{21}+\Gamma_{23})}\frac{2\mathrm{Re}[F_1]}{F_2|F_1|^2}\hat{a}^\dag\hat{a}^2,
\end{align}
\end{subequations}
where $F_2=\tilde{\gamma}_{34}+|\Omega_c|^2/\tilde{\gamma}_{24}$.

If $\hat{Q}(t)$ is an arbitrary combination of mode operators, the equation of motion for $\hat{Q}(t)$ is written from Eq.~\eqref{eq1:Rotated Hamiltonian} combining with the Lindblad operator Eq.~\eqref{eq1:Lindblad operator}
\begin{equation}\label{eq1:Q_operator_EOM}
\begin{split}
\dot{\hat{Q}}&=i\Delta\left[\hat{Q},\hat{a}^\dag\hat{a}\right]\\
     &\quad+g_1\left[\hat{Q},\hat{a}^\dag\right]\sum_{j=1}^{N}\hat{\sigma}_{12}^{j}-g_1\sum_{j=1}^{N}\hat{\sigma}_{21}^{j}\left[\hat{Q},\hat{a}\right]\\
     &\quad+g_2\left[\hat{Q},\hat{a}^\dag\right]\sum_{j=1}^{N}\hat{\sigma}_{34}^{j}-g_2\sum_{j=1}^{N}\hat{\sigma}_{43}^{j}\left[\hat{Q},\hat{a}\right]\\
     &\quad+\frac{\kappa}{2}\left[2\hat{a}^\dag\hat{Q}\hat{a}-\hat{a}^\dag\hat{a}\hat{Q}-\hat{Q}\hat{a}^\dag\hat{a}\right]\\
     &\quad+\sqrt{\kappa_{e,1}}\varepsilon_p\left[\hat{Q},\hat{a}^\dag-\hat{a}\right] \;.
\end{split}
\end{equation}
Substituting Eqs.~\eqref{eq1:third-order_sigma12}--\eqref{eq1:third-order_sigma34} into Eq.~\eqref{eq1:Q_operator_EOM} leads to
\begin{equation}\label{eq1:effective_Q_operator_EOM}
\begin{split}
\dot{\hat{Q}}&=i(\Delta-\delta\omega_\text{cav})\left[\hat{Q},\hat{a}^\dag\hat{a}\right]-i\eta\left[\hat{Q},\hat{a}^{\dag2}\hat{a}^2\right]\\
&\quad+\frac{\kappa_a^\mathrm{L}+\kappa}{2}\left[2\hat{a}^\dag\hat{Q}\hat{a}-\hat{a}^\dag\hat{a}\hat{Q}-\hat{Q}\hat{a}^\dag\hat{a}\right]\\
&\quad+\frac{\kappa_a^\mathrm{NL}}{2}\left[2\hat{a}^{\dag2}\hat{Q}\hat{a}^2-\hat{a}^{\dag2}\hat{a}^2\hat{Q}-\hat{Q}\hat{a}^{\dag2}\hat{a}^2\right]\\
&\quad+\sqrt{\kappa_{e,1}}\varepsilon_p\left[\hat{Q},\hat{a}^\dag-\hat{a}\right],
\end{split}
\end{equation}
where
\begin{subequations}
\begin{align}
\label{eq1:Re_chi1}\delta\omega_\text{cav}&=-g_1^2\sum_{j=1}^{N}\frac{\mathrm{Im}[F_1]}{|F_1|^2},\\
\kappa_a^\mathrm{L}&=2g_1^2\sum_{j=1}^{N}\frac{\mathrm{Re}[F_1]}{|F_1|^2},\\
\kappa_a^\mathrm{NL}&=4g_1^2\sum_{j=1}^{N}\left[-g_1^2Y_1\frac{\mathrm{Re}[F_1]^2}{|F_1|^4}+g_2^2Y_2\frac{\mathrm{Re}[F_2]\mathrm{Re}[F_1]}{|F_2|^2|F_1|^2}\right],\\
\label{eq1:Re_chi3}\eta&=2g_1^2\sum_{j=1}^{N}\left[g_1^2Y_1\frac{\mathrm{Im}[F_1]\mathrm{Re}[F_1]}{|F_1|^4}-g_2^2Y_2\frac{\mathrm{Im}[F_2]\mathrm{Re}[F_1]}{|F_2|^2|F_1|^2}\right],
\end{align}
\end{subequations}
with $Y_1=(\Gamma_{23}+2\Gamma_{31})/(\Gamma_{31}(\Gamma_{21}+\Gamma_{23}))$ and $Y_2=\Gamma_{23}/(\Gamma_{31}(\Gamma_{21}+\Gamma_{23}))$.

We derive the effective Hamiltonian of the system from Eq. \eqref{eq1:effective_Q_operator_EOM} to be
\begin{equation}\label{eq1:effective Hamiltonian}
\hat{H}=(-\Delta+\delta\omega_\text{cav})\hat{a}^\dag\hat{a}+\eta\hat{a}^{\dag2}\hat{a}^2+i\sqrt{\kappa_{e,1}}\varepsilon_p\left(\hat{a}^\dag-\hat{a}\right),
\end{equation}
and all the decay processes of this system, including the additional one-photon and two-photon absorption processes induced by atomic effects, are given by the effective Lindblad operator
\begin{equation}\label{eq1:effective Lindblad operator}
\begin{split}
\hat{\mathcal{L}}\hat{Q}&=\frac{\kappa_a^\mathrm{L}+\kappa}{2}\left[2\hat{a}^\dag\hat{Q}\hat{a}-\hat{a}^\dag\hat{a}\hat{Q}-\hat{Q}\hat{a}^\dag\hat{a}\right]+\\
&\quad\frac{\kappa_a^\mathrm{NL}}{2}\left[2\hat{a}^{\dag2}\hat{Q}\hat{a}^2-\hat{a}^{\dag2}\hat{a}^2\hat{Q}-\hat{Q}\hat{a}^{\dag2}\hat{a}^2\right].
\end{split}
\end{equation}
It is shown from Eq.~\eqref{eq1:effective Hamiltonian} that the cavity atomic system is now simulated as a cavity with a dispersive refractive index and a dispersive Kerr nonlinearity. The dispersive refractive index is indicated from the first term in Eq.~\eqref{eq1:effective Hamiltonian} where the resonance frequency of the equivalent cavity dispersively changes versus input field frequency. The resonance frequency of the cavity with the dispersive refractive index is $\omega'_\mathrm{cav}=q2\pi c/n(\omega'_\mathrm{cav})L=q2\pi c/\sqrt{1+(l_m/L)\mathrm{Re}[\chi^{(1)}](\omega'_\mathrm{cav})}L$, where $L$ is the round-trip length of intracavity photons, $l_m$ is the length traveled by a photon in the atomic medium in a round trip and $\chi^{(1)}$ is the linear susceptibility of the atoms. Then the resonance frequency of the equivalent dispersive cavity can be approximated as $\omega'_\mathrm{cav}\approx q2\pi c/L[1-(l_m/2L)\mathrm{Re}[\chi^{(1)}](\omega'_\mathrm{cav})]\equiv\omega_\mathrm{cav}+\delta\omega_\mathrm{cav}(\omega'_\mathrm{cav})$~\cite{Lasers}, in which $\omega_\mathrm{cav}\equiv q2\pi c/L$ is the bare cavity resonance and $\delta\omega_\mathrm{cav}(\omega'_\mathrm{cav})$ denotes the pulling of the bare cavity resonance which is induced by the atomic phase-shift effects.

We first consider the case that the equivalent dispersive cavity is resonantly driven, i.e., $\Delta=\delta\omega_\text{cav}(\omega'_\mathrm{cav})$ ($\omega_p=\omega_\text{cav}+\delta\omega_\text{cav}(\omega'_\mathrm{cav})\equiv\omega'_\text{cav}$), for which the related detunings are reduced to $\Delta_{21}=\omega_{21}-\omega'_\text{cav}$ and $\Delta_{43}=\omega_{43}-\omega'_\text{cav}$, then $F_1$ and $F_2$ in Eqs. \eqref{eq1:Re_chi1}--\eqref{eq1:Re_chi3} are reduced to functions of $\omega'_\text{cav}$ in the forms of
\begin{subequations}
\begin{align}
\label{eq1:F1ResonantCase}F_1&=i(\omega_{21}-\omega'_\text{cav})+\gamma_{12}+\frac{|\Omega_c|^2}{i(\omega_{21}-\omega'_\text{cav}-\Delta_{23})+\gamma_{13}},\\
\label{eq1:F2ResonantCase}F_2&=i(\omega_{43}-\omega'_\text{cav})+\gamma_{34}+\frac{|\Omega_c|^2}{i(\omega_{43}-\omega'_\text{cav}-\Delta_{23})+\gamma_{24}}.
\end{align}
\end{subequations}
Therefore the shift of the cavity resonance, the atomic one-photon absorption and the two-photon nonlinearities are determined by the detunings $\Delta_{21}$ and $\Delta_{43}$, or equivalently, the frequency $\omega^\prime_\mathrm{cav}$ of the actually resonant intra-cavity field.

For a slightly off-resonant frequency component in the probe laser, the effective cavity resonance is dispersively shifted, thus the detuning in the first term of Eq.~\eqref{eq1:effective Hamiltonian} is modified as $-\Delta'+\mathrm{d}[\delta\omega_\text{cav}](\Delta')$, in which we denote $\Delta^\prime\equiv\Delta-\delta\omega_\text{cav}(\omega'_\text{cav})\equiv\omega_p-\omega'_\text{cav}$, and $\mathrm{d}[\delta\omega_\text{cav}](\Delta')\equiv\delta\omega_\text{cav}(\omega_p)-\delta\omega_\text{cav}(\omega'_\text{cav})$ is the dispersive shift of the effective cavity resonance. Then the Hamiltonian Eq. \eqref{eq1:effective Hamiltonian} takes the form
\begin{equation}\label{eq1:probe_scanning Hamiltonian}
\hat{H}=\left(-\Delta'+\mathrm{d}[\delta\omega_\text{cav}](\Delta')\right)\hat{a}^\dag\hat{a}+\eta\hat{a}^{\dag2}\hat{a}^2+i\sqrt{\kappa_{e,1}}\varepsilon_p\left(\hat{a}^\dag-\hat{a}\right),
\end{equation}
for the scanning probe field. In the response to the scanning probe field, $\mathrm{d}[\delta\omega_\text{cav}](\Delta')$, $\eta$, $\kappa_a^\mathrm{L}$ and $\kappa_a^\mathrm{NL}$ change versus the detuning $\Delta'$ through $F_1(\Delta')$ and $F_2(\Delta')$, since the related detunings in $F_1$ and $F_2$ become $\Delta_{21}=\omega_{21}-\omega'_\text{cav}-\Delta'$, and $\Delta_{43}=\omega_{43}-\omega'_\text{cav}-\Delta'$, thus $F_1$ and $F_2$ are determined only by $\Delta'$.

\section{Results}\label{sec 2: results}
\begin{figure}
\centering
\includegraphics[width=0.64\linewidth]{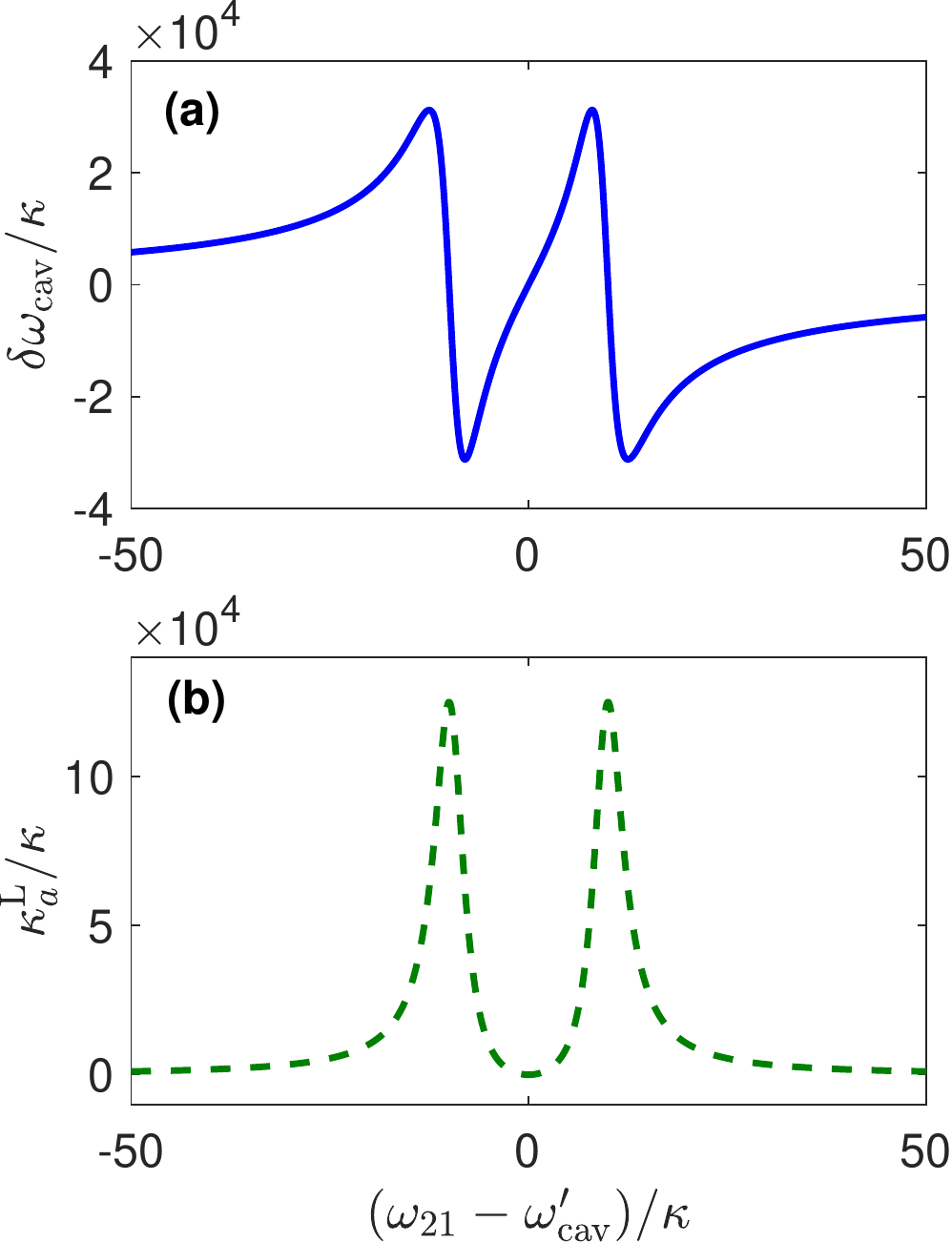}
\caption{(a) The pulling of the resonance frequency $\delta\omega_\mathrm{cav}$ and (b) the one-photon absorption rate $\kappa_a^\mathrm{L}$ as a function of the pulled cavity resonance frequency $\omega'_\text{cav}$, under the condition that the coupling laser is resonant with the $|3\rangle\leftrightarrow|2\rangle$ transition. We use $N=12.5\times10^6$, $g_1/\kappa=0.15$, $\Omega_c/\kappa=10$, $\Gamma_{31}/\kappa=10^{-5}$, $\Gamma_{21}/\kappa=\Gamma_{23}/\kappa=4.5$, $\omega_p=\omega'_\text{cav}$ and $\Delta_{23}=0$.}
\label{fig:Fig2}
\end{figure}

As proposed by A. Imamoglu \textit{et} \textit{al}.~\cite{GiantKerrNonlinearityEIT,StronglyInteractingPhotons}, a large kerr nonlinearity based on the electromagnetically induced transparency (EIT) have several advantages, including atomic one-photon loss elimination and vanishing pulling of the resonance frequency. If the cavity field at the resonance frequency $\omega_\mathrm{cav}$ and the coupling laser at $\omega_c$ are at two-photon resonance with the $|1\rangle\leftrightarrow|3\rangle$ transition, the transparency or a dark resonance is created at the cavity-mode frequency. In Fig. \ref{fig:Fig2}, the dispersive pulling of the bare cavity resonance $\delta\omega_\mathrm{cav}(\omega'_\text{cav})$ and the one-photon atomic absorption rate $\kappa_a^\mathrm{L}$ is plotted versus the effective resonance frequency $\omega'_\text{cav}$, under the condition that the coupling field is resonant with the $|3\rangle\leftrightarrow|2\rangle$ transition. It is verified from Fig.~\ref{fig:Fig2}(b) that the atomic one-photon absorption rate is negligibly small (about $0.02812\kappa$) when the intracavity field with frequency $\omega'_\text{cav}$ is also resonant with the $|1\rangle\leftrightarrow|2\rangle$ transition. Meanwhile, the pulling of the bare cavity resonance vanishes, as shown in Fig. \ref{fig:Fig2}(a), which indicates a cancelled linear atomic polarization of the $|1\rangle\leftrightarrow|2\rangle$ transition. Since the EIT scheme has these advantages, we use the two-photon resonance condition in the following context to eliminate linear atomic polarization and preserve third-order nonlinear polarization simultaneously, so as to avoid pulling the bare cavity resonance frequency and obtain a negligible one-photon absorption.

\begin{figure}
\centering
\includegraphics[width=0.64\linewidth]{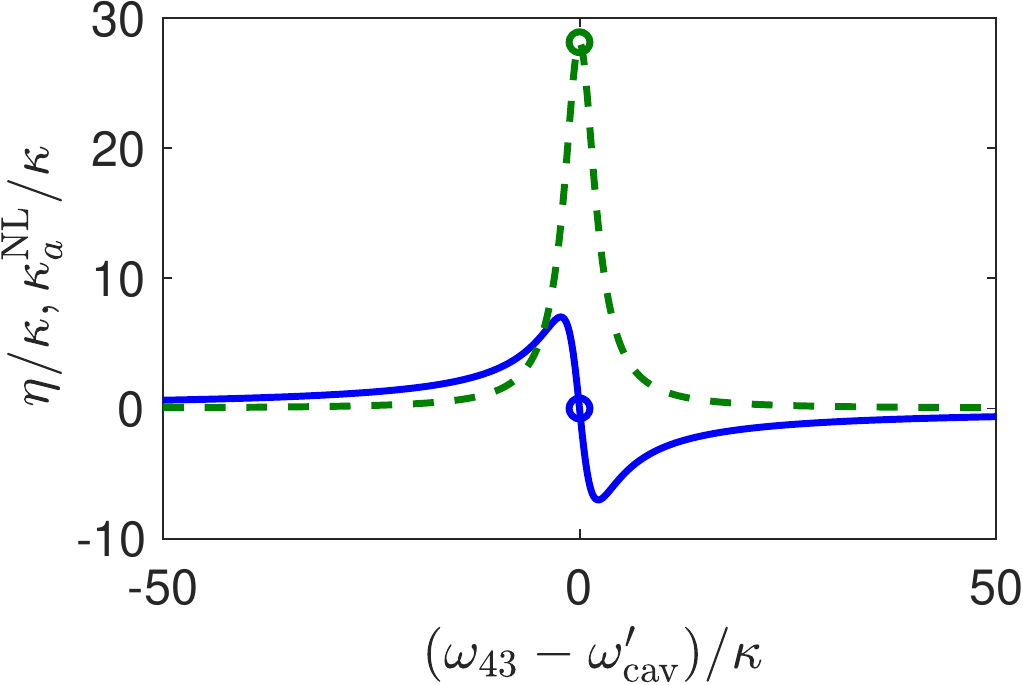}
\caption{The Kerr nonlinear coefficient $\eta$ (solid line) and the two-photon absorption rate $\kappa_a^\mathrm{NL}$ (dashed line) versus the detuning between the $|3\rangle\leftrightarrow|4\rangle$ transition and the pulled resonance frequency $\omega'_\text{cav}$ under the two-photon resonance condition. The parameters are taken as $N=12.5\times10^6$, $g_1/\kappa=g_2/\kappa=0.15$, $\Omega_c/\kappa=10$, $\Gamma_{31}/\kappa=10^{-5}$, $\Gamma_{21}/\kappa=\Gamma_{23}/\kappa=\Gamma_{43}/\kappa=4.5$, $(\omega_{21}-\omega'_\text{cav})/\kappa=\Delta_{23}/\kappa=4560$.}
\label{fig:Fig3}
\end{figure}

In Fig.~\ref{fig:Fig3}, the Kerr nonlinear coefficient $\eta$ and the two-photon absorption rate $\kappa_a^\mathrm{NL}$ are plotted as functions of the detuning between the $|3\rangle\leftrightarrow|4\rangle$ transition and the pulled resonance frequency $\omega'_\text{cav}$ under the two-photon resonance condition. We set the detuning between the $|1\rangle\leftrightarrow|2\rangle$ and the $|3\rangle\leftrightarrow|4\rangle$ transitions to be $4560\kappa$, the two-photon resonance condition is taken as $\omega_{21}-\omega'_\text{cav}=\Delta_{23}=4560\kappa$ to make the cavity field nearly resonantly couples to the $|3\rangle\leftrightarrow|4\rangle$ transition. To realize antibunched photons via the two-photon absorption instead of a conservative Kerr nonlinearity, we want a nearly imaginary $\chi^{(3)}$ nonlinearity. Therefore we choose the parametric point corresponding to $\eta/\kappa\approx0$, marked in a blue circle with the detuning $\omega_{43}-\omega'_\text{cav}=-0.0219\kappa$ and $\eta/\kappa\sim10^{-4}$, as shown in Fig.~\ref{fig:Fig3}. The relevant nonlinear dissipation rate is $\kappa_a^\mathrm{NL}/\kappa=28.12$, which is marked in a green circle. The related one-photon absorption rate and the pulling of the resonance frequency are $\kappa_a^\mathrm{L}/\kappa=0.02812$ and $\delta\omega_\mathrm{cav}(\omega'_\text{cav})/\kappa\sim10^{-6}$. We take this parameter configuration that $\omega_{21}-\omega'_\text{cav}=\Delta_{23}=4560\kappa$ and $\omega_{43}-\omega'_\text{cav}=-0.0219\kappa$ in the following context, which leads to a resonantly driven cavity with a negligible linear atomic effect and a large, purely absorptive two-photon nonlinearity. In a brief summary, one-photon state propagates without loss as in vacuum, as the atomic medium is reduced to its $\Lambda$-type subsystem for one-photon excitation, where EIT eliminates one-photon absorption. Whereas for the simultaneous arrival of two or more photons, the complete N-type level structure works by which the two-photon absorption is switched on.

\begin{figure}
\centering
\includegraphics[width=0.62\linewidth]{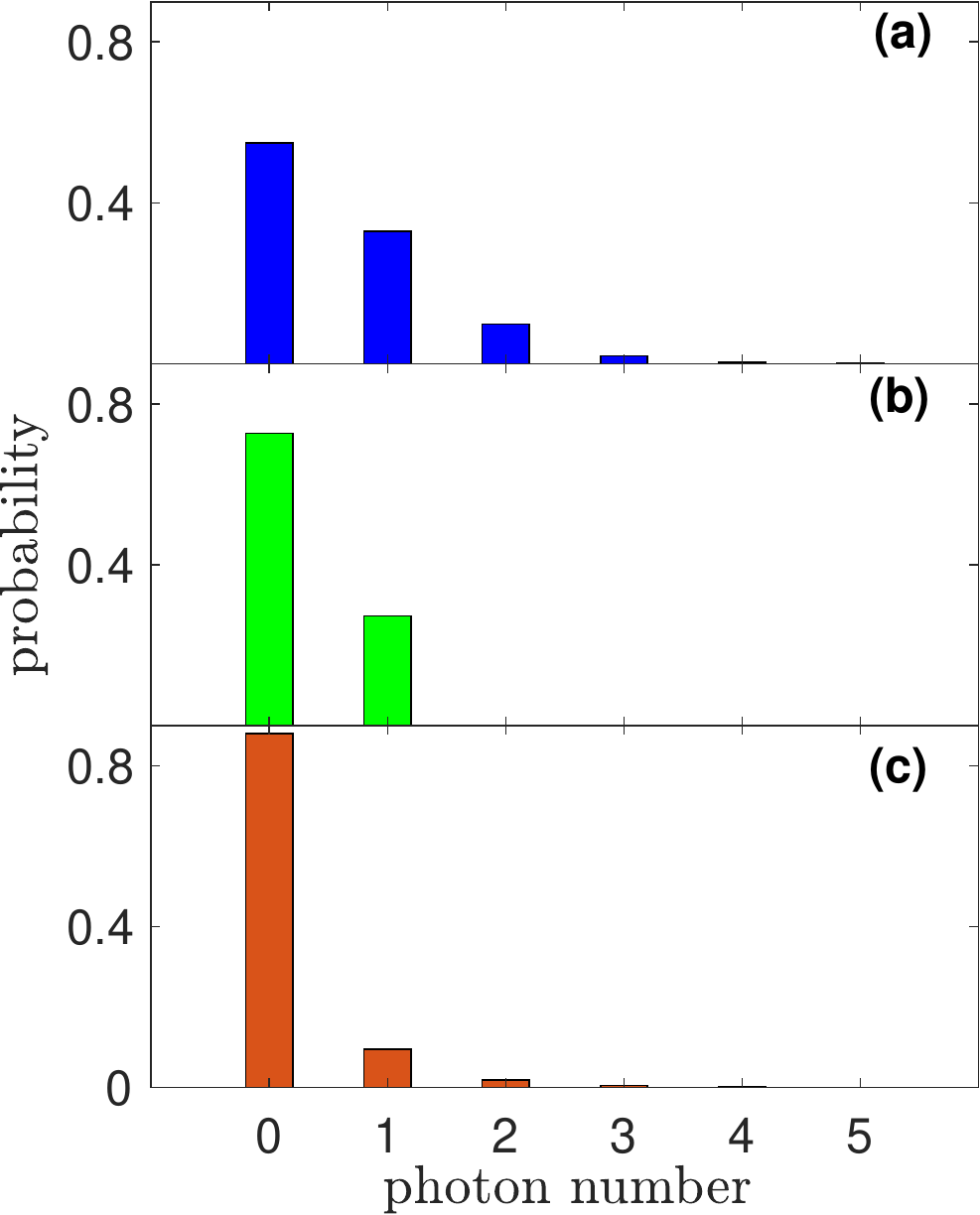}
\caption{In the nonlinearly dissipative cavity resonantly driven by the coherent field, the Fock-state probabilities of (a) the incident mode, (b) the transmitted mode and (c) the reflected mode. The parameters are taken as $\kappa_a^\mathrm{L}=0$, $\kappa_{e1}=\kappa_{e2}=0.5\kappa$ and $\kappa_i=0$; $\eta=0$ and $\kappa_a^\mathrm{NL}/\kappa=28$.}
\label{fig:Fig4}
\end{figure}

In what follows we verify the mechanism of the PB induced by a considerable two-photon absorption process. We study the reflected and transmitted fields of the nonlinearly dissipative cavity resonantly driven by a coherent input, with the incident and reflected radiation being quantized by the quantum cascade method \cite{QuantumTrajectory-CascadedOpenSystem,QuantumCascade1,QuantumCascade2} which enabled the investigation of their statistics (see Appendix \ref{appendix B} for details). We use the creation operators $\hat{d}^\dag$, $\hat{a}^\dag$ and $\hat{c}^\dag$ to define the generation of photons in the incident mode, transmitted mode (intracavity mode) and reflected mode respectively. The Fock-state probabilities of the cavity mode (incident and reflected modes), $P_{n_a}(t)=\langle n_a|\rho(t)|n_a\rangle$ ($P_{n_i}(t)=\langle n_i|\rho(t)|n_i\rangle,\,i=d,c$), are plotted in Fig. \ref{fig:Fig4}. It is shown in Fig. \ref{fig:Fig4}(a) that the Fock-state components in the state of the incident field satisfies the Poisson distribution with mean photon number $\bar{n}=0.6$. As shown in Fig.~\ref{fig:Fig4}(b), multi-photon components in the coherent input can not exist in the cavity because they dissipate to vacuum or one-photon states by the two-photon absorption channel behind the input port. Very few of them decay out of the cavity at the input port as a mixture of single-photon and multi-photon states, as shown in Fig.~\ref{fig:Fig4}(c). While the incident single photons are transmitted without loss through the cavity behind the input port with a portion of themselves decaying at the input channel, as shown in Figs.~\ref{fig:Fig4}(b) and \ref{fig:Fig4}(c). This phenomenon is consistent with the physical picture illustrated in the previous context by Fig. \ref{fig:Fig1}(c).

\begin{figure}
\centering
\includegraphics[width=0.72\linewidth]{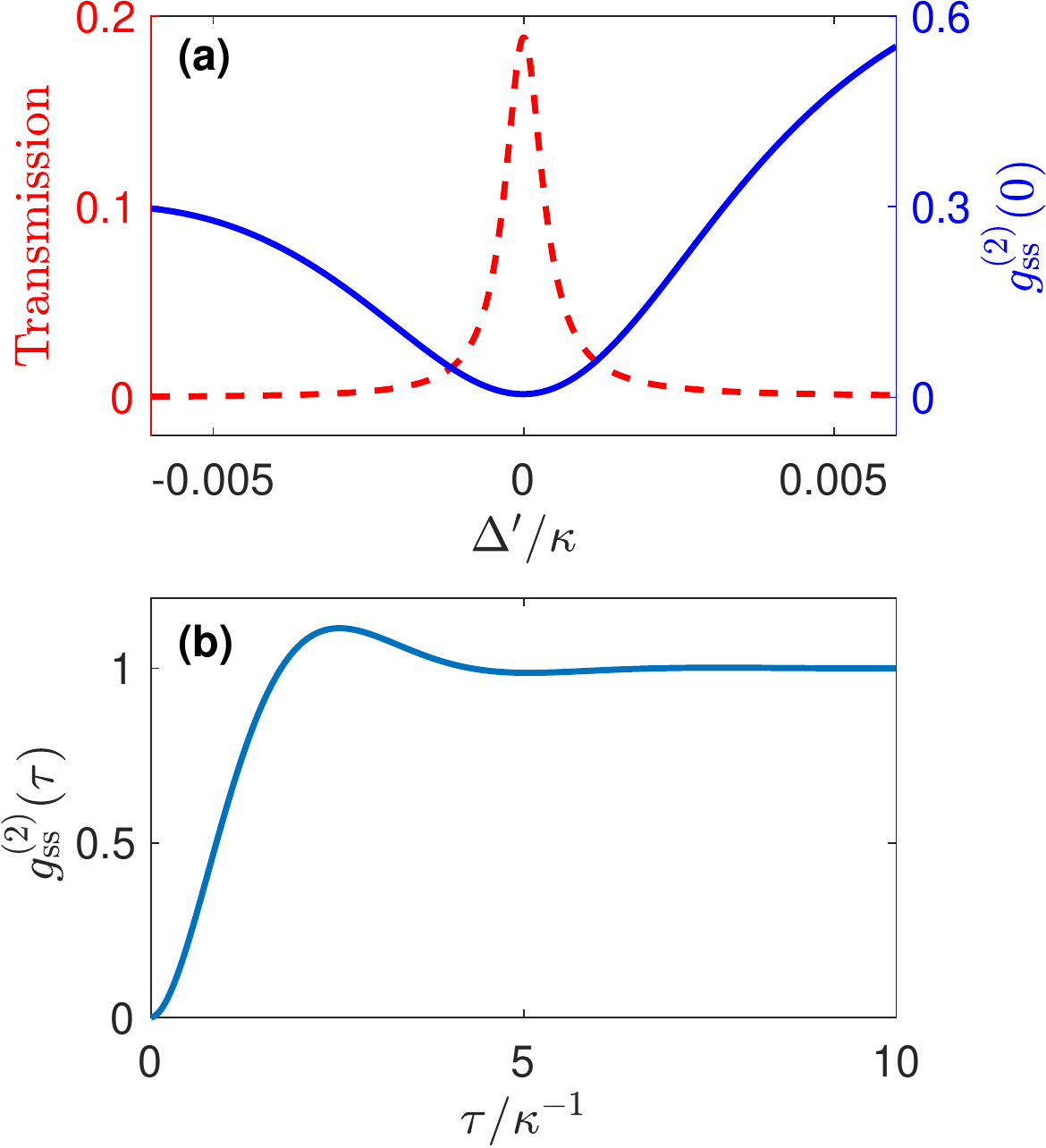}
\caption{(a) Transmission of the cavity field (dashed line) and the equal-time second-order correlation function at the steady state $g^{(2)}_\text{ss}(0)$ (solid line) versus the detuning $\Delta'$ between probe laser and the modified resonance, (b) the steady-state second-order correlation function with delay time $\tau$ versus the delay time. The parameters are taken as $N=12.5\times10^6$, $g_1/\kappa=g_2/\kappa=0.15$, $\Omega_c/\kappa=10$, $\Gamma_{31}/\kappa=10^{-5}$, $\Gamma_{21}/\kappa=\Gamma_{23}/\kappa=\Gamma_{43}/\kappa=4.5$, $(\omega_{21}-\omega'_\text{cav})/\kappa=\Delta_{23}/\kappa=4560$, $(\omega_{43}-\omega'_\text{cav})/\kappa=-0.0219$, which leads to $\kappa_a^\mathrm{L}/\kappa=0.02812$, $\eta/\kappa\sim10^{-4}$ and $\kappa_a^\mathrm{NL}/\kappa=28.12$ for resonant probe field. The remaining parameters are $\kappa_{e1}/\kappa=\kappa_{e2}/\kappa=0.45$ and $\kappa_i/\kappa=0.1$.}
\label{fig:Fig5}
\end{figure}

Then we explore the properties of the transmission spectrum and the steady-state second-order correlation function when the probe light scans around the modified resonance frequency. The transmission of photons is defined as $T=\langle\hat{a}_\text{out}^\dag\hat{a}_\text{out}\rangle/|\varepsilon_p|^2$, where $\hat{a}_\text{out}=\sqrt{\kappa_{e2}}\hat{a}$ is the annihilation operator of the transmitted field. It is shown in Fig.~\ref{fig:Fig5}(a) that the cavity linewidth of the transmission is significantly narrowed, which can be illustrated by the dispersive refractive index. As known from Eq.~\eqref{eq1:probe_scanning Hamiltonian}, although the detuning between the probe laser and the effective cavity resonance is $\Delta^\prime$, the actual detuning is modified by the dispersive shift of the effective cavity resonance versus the scanning probe frequency, which is denoted as $\mathrm{d}[\delta\omega_\text{cav}]\equiv\delta\omega_\text{cav}(\omega_p)-\delta\omega_\text{cav}(\omega'_\text{cav})$ in Eq.~\eqref{eq1:probe_scanning Hamiltonian}. As shown in Fig.~\ref{fig:Fig6}(a) of Appendix \ref{appendix A}, the dispersive shift of the effective cavity resonance is very sharp versus scanning probe frequency, which switches a slightly off-resonant frequency component out of the cavity resonance, leading to the significant cavity linewidth narrowing (see Appendix \ref{appendix A} for detailed derivation). For the probe field scanning through the narrowed cavity linewidth, the Kerr nonlinear coefficient and the atomic one-photon absorption rate are always negligible compared with the cavity decay rate, while the two-photon absorption rate remains large, as shown in Fig.~\ref{fig:Fig6}(b)--\ref{fig:Fig6}(d) of Appendix \ref{appendix A}. Namely, the N-type quantum system can be modeled as a cavity with a dispersive refractive index which has a large two-photon absorption process in the response to the scanning probe field. The equal-time second-order correlation function at the steady state, which is defined as $g^{(2)}_\text{ss}(0)=\lim_{t\rightarrow\infty}\langle\hat{a}^\dag\hat{a}^\dag\hat{a}\hat{a}\rangle(t)/\langle\hat{a}^\dag\hat{a}\rangle^2(t)$, measures the variance of the photon-number distribution of the cavity field and represents the probability of two-photon transmission. It's shown in Fig.~\ref{fig:Fig5}(a) that $g^{(2)}_\text{ss}(0)$ reaches its minimum about $0.005$ at the resonance and $g^{(2)}_\text{ss}(0)<0.05$ within the narrowed cavity linewidth, indicating that the transmitted photons are well anti-bunched and sub-Poissonian when the probe laser drives the nonlinearly dissipative cavity within the linewidth. The steady-state second-order correlation function with time delay $\tau$, which gives the joint probability of detecting a second photon at time $\tau$ given a detection event that starts from the steady state at time $t=0$, is defined as $g^{(2)}_\text{ss}(\tau)=\lim_{t\rightarrow\infty}\langle\hat{a}^\dag(t)\hat{a}^\dag(t+\tau)\hat{a}(t+\tau)\hat{a}(t)\rangle/\langle\hat{a}^\dag\hat{a}\rangle^2(t)$. In Fig.~\ref{fig:Fig5}(b), $g^{(2)}_\text{ss}(\tau)$ is plotted versus the delay time $\tau$ for the resonantly driven system. It's shown that $g^{(2)}_\text{ss}(\tau)$ increases with the delay time, indicating photons are more likely to arrive separated in time, which verifies that the sub-Poissonian photons transmitted within the cavity linewidth are indeed anti-bunched.

\section{Experimental implementation}\label{sec 3: implementation}
Our scheme can be implemented with a setup of a Fabry-P\'{e}rot cavity with a $\prescript{87}{}{\mathrm{Rb}}$ cell placed inside, as depicted in Fig. 1(a). The two mirrors have the same reflectivity of $99\%$. End faces of the atomic cell are coated with $99.9\%$ antireflection layers for the cavity mode. We use the 0.4-m-long Fabry-Perot cavity. The cavity internal losses is calculated to be $\kappa_i=2\pi\times0.12\,\mathrm{MHz}$ and the external loss rates at the two ports are $\kappa_{e1}=\kappa_{e2}=2\pi\times0.6\,\mathrm{MHz}$, yielding $\kappa=2\pi\times1.32\,\mathrm{MHz}$. We exploit the D1 line of $\prescript{87}{}{\mathrm{Rb}}$ atom to realize the N-type configuration with $|1\rangle=|5\prescript{2}{}{S}_{1/2},F=1,m_F=-1\rangle$, $|3\rangle=|5\prescript{2}{}{S}_{1/2},F=2,m_F=-2\rangle$, $|2\rangle=|5\prescript{2}{}{P}_{1/2},F'=1,m'_F=-1\rangle$, $|4\rangle=|5\prescript{2}{}{P}_{1/2},F'=2,m'_F=-2\rangle$. The coupling laser is left circularly polarized and the probe laser is linearly polarized. The single-photon atomic coupling strength is calculated to be about $g_1=g_2=2\pi\times0.2\,\mathrm{MHz}$ for the mode volume of this cavity and the dipole matrix element of D1 transition. In the related hyperfine levels in the D1 line of $\prescript{87}{}{\mathrm{Rb}}$ atoms, the detuning between the 1--2 and the 3--4 transitions is about $2\pi\times6020\,\mathrm{MHz}$, thus we set the detuning between the $|1\rangle\leftrightarrow|2\rangle$ and $|3\rangle\leftrightarrow|4\rangle$ transitions to be $4560\kappa$ in the previous context. The decay rates of the hyperfine-level transitions in $\prescript{87}{}{\mathrm{Rb}}$ D1 line are about $2\pi\times6\,\mathrm{MHz}$.

\section{Discussion and conclusion}\label{sec 4: conclusion}
N-type systems are usually exploited to realize a giant Kerr nonlinearity owing to the cancelled linear susceptibility and the enhanced nonlinear susceptibility in EIT scheme. Subsequently, the PB effect induced by absorptive-free nonlinear dispersion has been extensively researched, the mechanism behind which is that the large nonlinear phase shifts of multi-photons enable an anharmonic eigen-energy level structure of photons. However, the dispersive Kerr nonlinearity is suppressed by the large detuning in the nonresonant case. Counterintuitively, we eliminate the Kerr nonlinearity but keep a significant two-photon absorption by selecting near-resonant optical nonlinear processes. In such configuration, we achieve deep PB. Our scheme for nonlinear dissipation induced PB is easier to implement and more efficient.

In summary, we have proposed a scheme to realize the deep PB effect by inducing the large nonlinear dissipation of an optical cavity with N-type atoms. In particular, a large and dominant two-photon absorption is achieved in the cavity by means of near-resonant nonlinear process, whereas the atomic one-photon absorption is suppressed to be vanishing. The deep PB is accessible within the linewidth of the N-type quantum system which is completely induced by the nonlinear dissipation, and a high transmission efficiency is shown simultaneously. Our proposal provides a potential protocol for the efficient PB and may open up a new route towards manipulation of single photons.


\section*{Acknowledgement}
This work is supported by the National Key R\&D Program of China (Grants No. 2019YFA0308700, No. 2019YFA0308704 and No. 2017YFA0303703), the National Natural Science Foundation of China (Grants No. 11874212, No. 11890704), the Program for Innovative Talents and Teams in Jiangsu (Grant No. JSSCTD202138), and the Excellent Research Program of Nanjing University (Grant No. ZYJH002).

\appendix
\section{cavity-linewidth narrowing}\label{appendix A}
\begin{figure}
\centering
\includegraphics[width=\linewidth]{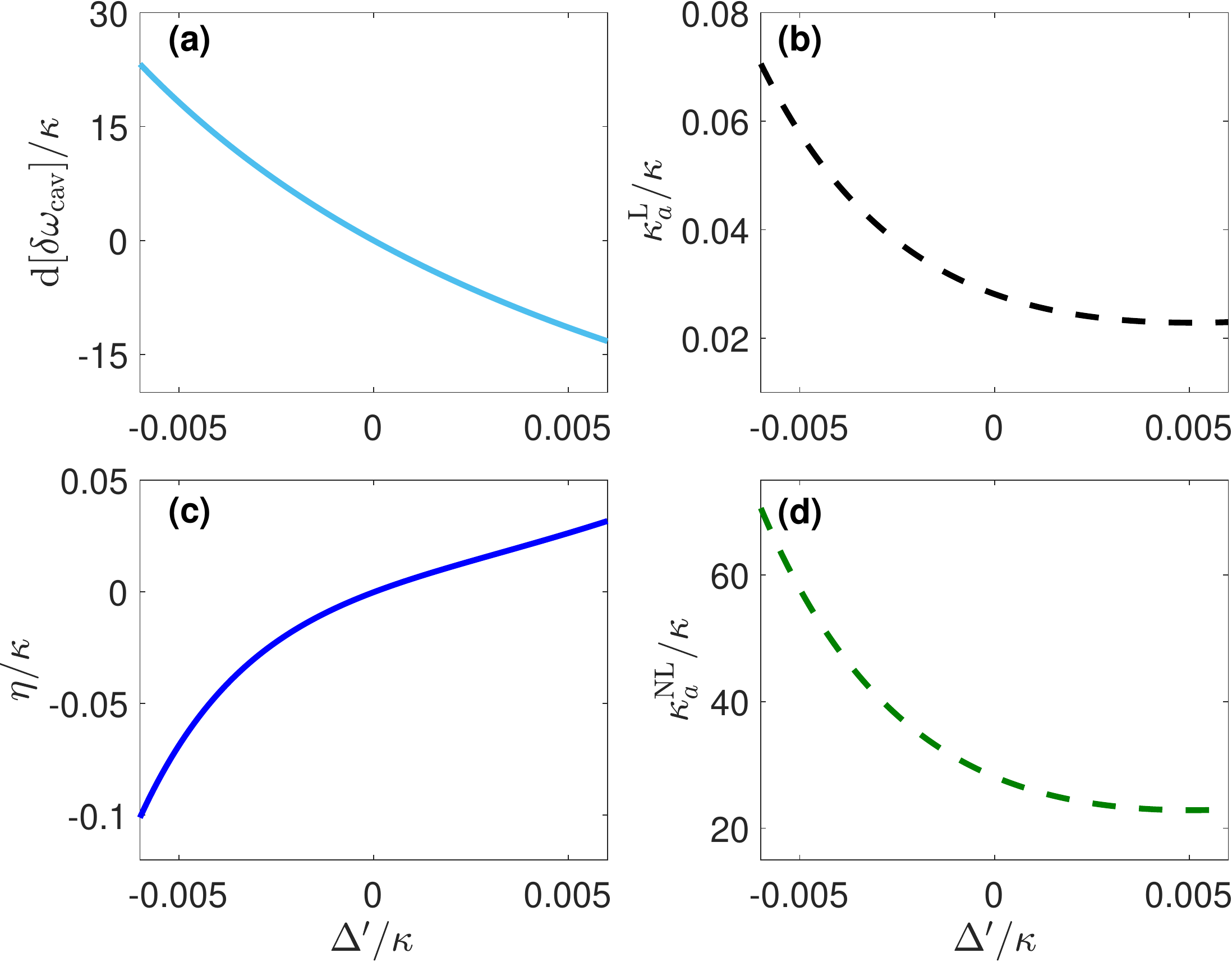}
\caption{(a) The dispersive shift of the effective cavity resonance, (b) the atomic one-photon absorption rate, (c) the Kerr nonlinear coefficient and (d) the atomic two-photon absorption rate versus the detuning $\Delta'$. All the parameters are the same as in Fig. \ref{fig:Fig5}.}
\label{fig:Fig6}
\end{figure}

The dispersive shift of the effective optical resonance, the Kerr nonlinear coefficient and the atomic one-photon and two-photon absorption rates vary versus the detuning $\Delta'$ as shown in Fig. \ref{fig:Fig6}. A slightly off-resonant probe frequency with the detuning $\Delta'=-0.005\kappa$ can see a large dispersive shift of the effective cavity resonance up to $\mathrm{d}[\delta\omega_\text{cav}]=18.22\kappa$, which switches itself out of the cavity resonance, leading to the significant cavity linewidth narrowing effect. The Kerr nonlinear coefficient and the atomic one-photon absorption rate are always negligibly small compared to cavity decay rate when the probe frequency scans through the narrowed linewidth, as shown in Figs. \ref{fig:Fig6}(b) and \ref{fig:Fig6}(c). Nevertheless, the two-photon absorption rate is always large within the linewidth, as shown in Fig. \ref{fig:Fig6}(d).

Then we briefly deduce the cavity linewidth narrowing effect to better understand the results in the main context. Suppose the atomic medium with the linear susceptibility $\chi(\omega)=\chi'(\omega)+i\chi''(\omega)$ and the length $l_m/2$ is placed in the Fabry-P\'{e}rot cavity of length $L/2$. Inside the atomic medium, the propagation constant of the radiation field is $\beta_m=\beta_0\sqrt{1+\chi'(\omega)}\approx\beta_0+\Delta\beta_m(\omega)$ where $\Delta\beta_m(\omega)=\beta_0\chi'(\omega)/2$ and $\beta_0$ is the propagation constant in the air. The round-trip phase shift $\phi(\omega)$ in the cavity is
\begin{equation}
\phi(\omega)=\frac{\omega}{c}L+\frac{\omega\chi'(\omega)}{2c}l_m.
\end{equation}
The complex amplitude of the circulating intracavity field is related to that of the incident field outside the cavity by
\begin{equation}
\tilde{E}_\text{circ}=it_1\tilde{E}_\text{inc}+\tilde{g}_\text{rt}(\omega)\tilde{E}_\text{circ},
\end{equation}
where $\tilde{g}_\text{rt}(\omega)=r_1r_2e^{-\alpha L-i(\omega L/c+\omega\chi'(\omega)l_m/2c)}$ with $r_1$, $t_1$, $r_2$ and $\alpha$ being the reflection and transmission coefficients of the input-port mirror, the reflection coefficient of the second mirror and the attenuation constant inside cavity respectively. The ratio of the complex amplitude of the circulating intracavity field to that of the incident field is given by
\begin{equation}
\frac{\tilde{E}_\text{circ}}{\tilde{E}_\text{inc}}=\frac{it_1}{1-\tilde{g}_\text{rt}(\omega)}.
\end{equation}
The spectrum of the intracavity field is
\begin{equation}
\begin{split}
T&=\frac{|t_1|^2}{|1-\tilde{g}_\text{rt}(\omega)|^2}\\
&=\frac{|t_1|^2}{(1-g_\text{rt})^2+4g_\text{rt}\sin^2(\omega L/2c+\omega\chi'(\omega)l_m/4c)},
\end{split}
\end{equation}
where $g_\text{rt}=|\tilde{g}_\text{rt}|$. The intracavity field intensity reduce to half its maximum value when the frequency satisfies
\begin{equation}
\frac{\omega L}{2c}+\frac{\omega\chi'(\omega)}{4c}l_m=q\pi\pm\arcsin(\frac{1-g_\text{rt}}{2\sqrt{g_\text{rt}}}),
\end{equation}
thus the frequencies corresponding to the half intracavity field intensity are
\begin{equation}
\begin{split}
\omega_\pm&=\frac{q2\pi c/L\pm2c/L\arcsin(\frac{1-g_\text{rt}}{2\sqrt{g_\text{rt}}})}{1+\frac{\chi'(\omega_\pm)l_m}{2L}}\\
&=\left[q\frac{2\pi c}{L}\pm\frac{2c}{L}\arcsin(\frac{1-g_\text{rt}}{2\sqrt{g_\text{rt}}})\right]\\
&\quad\times\left[1-\frac{l_m}{L}\frac{\chi'(\omega_\pm)}{2}\right],
\end{split}
\end{equation}
leading to
\begin{subequations}
\begin{align}
\omega_+&=(\omega_\text{cav}+\frac{\Delta\omega}{2})(1-\frac{l_m}{2L}\chi'(\omega_+)),\\
\omega_-&=(\omega_\text{cav}-\frac{\Delta\omega}{2})(1-\frac{l_m}{2L}\chi'(\omega_-)),
\end{align}
\end{subequations}
where $\omega_\text{cav}=q2\pi c/L$ is the bare cavity resonance frequency and $\Delta\omega=4c/L\arcsin((1-g_\text{rt})/2\sqrt{g_\text{rt}})$ is the linewidth of the bare cavity. Thus the modified cavity linewidth is obtained as
\begin{equation}
\begin{split}
\Delta\omega'&=\omega_+-\omega_-\\
&\approx-\frac{l_m}{2L}\omega_\text{cav}\chi'(\omega_+)+\frac{l_m}{2L}\omega_\text{cav}\chi'(\omega_-)+\Delta\omega\\
&<\Delta\omega
\end{split}
\end{equation}
In the above equation we use the condition that $\chi'(\omega_+)>0$, and $\chi'(\omega_-)<0$. The modified cavity linewidth is narrowed compared to the bare cavity linewidth.
\section{quantum cascade to quantize the input of the nonlinearly dissipative cavity}\label{appendix B}
\begin{figure}
\centering
\includegraphics[width=1\linewidth]{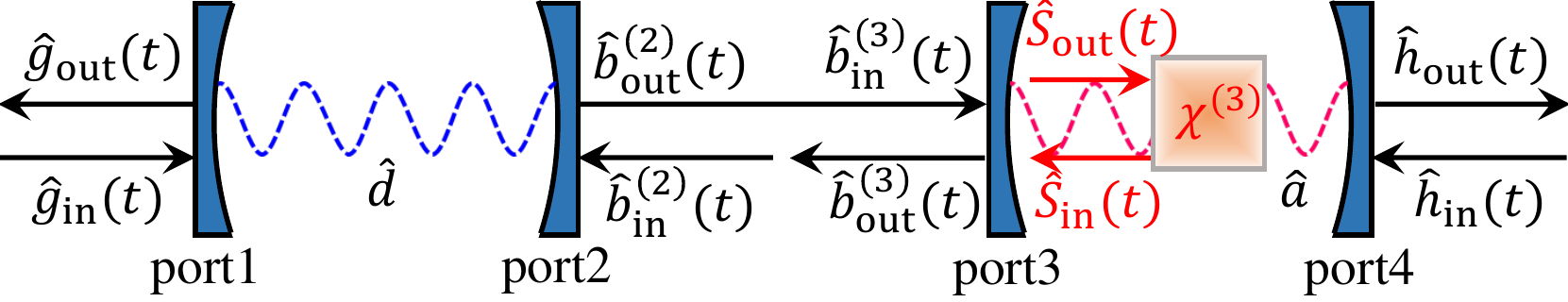}
\caption{The schematic diagram of the cascaded quantum system. The output of the first cavity at port 2 is guided to drive the second cavity at port 3. The second cavity contains atomic ensembles and thus experiences a nonlinear dissipation process.}
\label{fig:Fig7}
\end{figure}
To investigate the statistics of the reflected field of the nonlinearly dissipative cavity, the incident field should be quantized radiation. We consider a cascaded quantum system \cite{QuantumTrajectory-CascadedOpenSystem,QuantumCascade1,QuantumCascade2} which can be decomposed into two subsystems, as depicted in Fig. \ref{fig:Fig7}. The first subsystem is an empty cavity driven by the coherent light from port 1, whose fluorescent output from port 2 is fed into the input port of the second subsystem, namely the localized nonlinearly dissipative cavity. The Hamiltonian for the cascaded system is
\begin{equation}
\begin{split}
\hat{H}=&\hat{H}_\text{sys}+\int_{-\infty}^{\infty}d\omega\,\omega\hat{g}^\dag(\omega)\hat{g}(\omega)+\int_{-\infty}^{\infty}d\omega\,\omega\hat{b}^\dag(\omega)\hat{b}(\omega)\\
&+\int_{-\infty}^{\infty}d\omega\,\omega\hat{h}^\dag(\omega)\hat{h}(\omega)+\int_{-\infty}^{\infty}d\Omega\,\Omega\hat{S}^\dag(\Omega)\hat{S}(\Omega)\\
&+i\int_{-\infty}^{\infty}d\omega\,\gamma_1(\omega)\left\{\hat{g}^\dag(\omega)\hat{d}-\hat{d}^\dag\hat{g}(\omega)\right\}\\
&+i\int_{-\infty}^{\infty}d\omega\,\gamma_2(\omega)\left\{\hat{b}^\dag(\omega)\hat{d}-\hat{d}^\dag\hat{b}(\omega)\right\} \\
&+i\int_{-\infty}^{\infty}d\omega\,\gamma_3(\omega)\left\{\hat{b}^\dag(\omega)e^{-i\omega\tau}\hat{a}-\hat{a}^\dag\hat{b}(\omega)e^{i\omega\tau}\right\}\\
&+i\int_{-\infty}^{\infty}d\omega\,\gamma_4(\omega)\left\{\hat{h}^\dag(\omega)\hat{a}-\hat{a}^\dag\hat{h}(\omega)\right\}\\
&+i\int_{-\infty}^{\infty}d\Omega\,G(\Omega)\left\{\hat{S}^\dag(\Omega)\hat{a}^2-\hat{a}^{\dag2}\hat{S}(\Omega)\right\}.
\end{split}
\end{equation}
We shall adapt the input-output theory. We now introduce the first Markov approximation, that the coupling constant is independent of frequency, that is
\begin{equation}
\begin{split}
\gamma_1(\omega)&=\sqrt{\kappa_{d1}/2\pi},\quad\gamma_2(\omega)=\sqrt{\kappa_{d2}/2\pi},\\
\gamma_3(\omega)&=\sqrt{\kappa_{e1}/2\pi},\quad\gamma_4(\omega)=\sqrt{\kappa_{e2}/2\pi},\\
G(\Omega)&=\sqrt{\kappa_a^\mathrm{NL}/2\pi}.
\end{split}
\end{equation}
Then the quantum Langevin equation for an arbitrary operator of the system $\hat{q}$ is derived as
\begin{equation}
\begin{split}
\dot{\hat{q}}=&-i\left[\hat{q}, \hat{H}_\text{sys}\right]-\left[\hat{q}, \hat{d}^\dag\right]\left\{\frac{\kappa_d}{2}\hat{d}+\sqrt{\kappa_{d1}}\hat{g}_\text{in}(t)+\sqrt{\kappa_{d2}}\hat{b}_\text{in}(t)\right\}\\
&+\left\{\frac{\kappa_d}{2}\hat{d}^\dag+\sqrt{\kappa_{d1}}\hat{g}^\dag_\text{in}(t)+\sqrt{\kappa_{d2}}\hat{b}_\text{in}^\dag(t)\right\}\left[\hat{q}, \hat{d}\right]\\
&-\left[\hat{q}, \hat{a}^{\dag2}\right]\left\{\frac{\kappa_a^\mathrm{NL}}{2}\hat{a}^2+\sqrt{\kappa_a^\mathrm{NL}}\hat{S}_\text{in}(t)\right\}\\
&+\left\{\frac{\kappa_a^\mathrm{NL}}{2}\hat{a}^{\dag2}+\sqrt{\kappa_a^\mathrm{NL}}\hat{S}^\dag_\text{in}(t)\right\}\left[\hat{q}, \hat{a}^2\right]\\
&-\left[\hat{q}, \hat{a}^\dag\right]\left\{\frac{\kappa_a}{2}\hat{a}+\sqrt{\kappa_{d2}\kappa_{e1}}\hat{d}(t-\tau)+\sqrt{\kappa_{e1}}\hat{b}_\text{in}(t-\tau)\right\}\\
&+\left\{\frac{\kappa_a}{2}\hat{a}^\dag+\sqrt{\kappa_{d2}\kappa_{e1}}\hat{d}^\dag(t-\tau)+\sqrt{\kappa_{e1}}\hat{b}_\text{in}^\dag(t-\tau)\right\}\left[\hat{q}, \hat{a}\right]\\
&-\sqrt{\kappa_{e2}}\left[\hat{q}, \hat{a}^\dag\right]\hat{h}_\text{in}(t)+\sqrt{\kappa_{e2}}\hat{h}_\text{in}^\dag(t)\left[\hat{q}, \hat{a}\right],
\end{split}
\end{equation}
in which $\tau$ is the propagation time for light to travel from the source cavity to the target cavity, with the input-output relations for the cascaded system at the internally-connected ports 2 and 3 given by
\begin{subequations}
\begin{align}
\hat{b}^{(2)}_\text{out}(t)&=\hat{b}_\text{in}(t)+\sqrt{\kappa_{d2}}\hat{d}(t),\\
\hat{b}^{(3)}_\text{out}(t)&=\hat{b}_\text{in}(t-\tau)+\sqrt{\kappa_{d2}}\hat{d}(t-\tau)+\sqrt{\kappa_{e1}}\hat{a}(t).
\end{align}
\end{subequations}
Taking the limit of negligible propagation time between localized components, i.e. $\tau\rightarrow0$, and considering simultaneously that $\hat{b}_\text{in}(t)$ and $\hat{h}_\text{in}(t)$ are vacuum fluctuations and $\hat{g}_\text{in}(t)=\alpha(t)$ is a coherent input, it follows that the master equation for the density operator $\hat{\rho}(t)$ can be derived by setting $\langle\dot{\hat{q}}\hat{\rho}\rangle\equiv\langle\hat{q}\dot{\hat{\rho}}\rangle$. The master equation takes the form
\begin{equation}
\begin{split}
\dot{\hat{\rho}}=&-i\left[\hat{H}_\text{sys},\hat{\rho}\right]-\frac{\kappa_d}{2}\left[\hat{d}^\dag,\hat{d}\hat{\rho}\right]+\frac{\kappa_d}{2}\left[\hat{d},\hat{\rho}\hat{d}^\dag\right]\\
&-\frac{\kappa_a}{2}\left[\hat{a}^\dag,\hat{a}\hat{\rho}\right]+\frac{\kappa_a}{2}\left[\hat{a},\hat{\rho}\hat{a}^\dag\right]\\
&-\sqrt{\kappa_{d2}\kappa_{e1}}\left[\hat{a}^\dag,\hat{d}\hat{\rho}\right]+\sqrt{\kappa_{d2}\kappa_{e1}}\left[\hat{a},\hat{\rho}\hat{d}^\dag\right]\\
&-\frac{\kappa_a^\mathrm{NL}}{2}\left[\hat{a}^{\dag2},\hat{a}^2\hat{\rho}\right]+\frac{\kappa_a^\mathrm{NL}}{2}\left[\hat{a}^2,\hat{\rho}\hat{a}^{\dag2}\right]\\
&-\sqrt{\kappa_{d1}}\left[\alpha(t)\hat{d}^\dag-\alpha^\ast(t)\hat{d},\hat{\rho}\right],
\end{split}
\end{equation}
where the Lindblad operator in the third line accounts for the cascaded coupling, that is, the output from the source cavity can be connected to the input of the target cavity without there being a corresponding scattering from the target cavity back into the source cavity.

\begin{figure}
\centering
\includegraphics[width=0.62\linewidth]{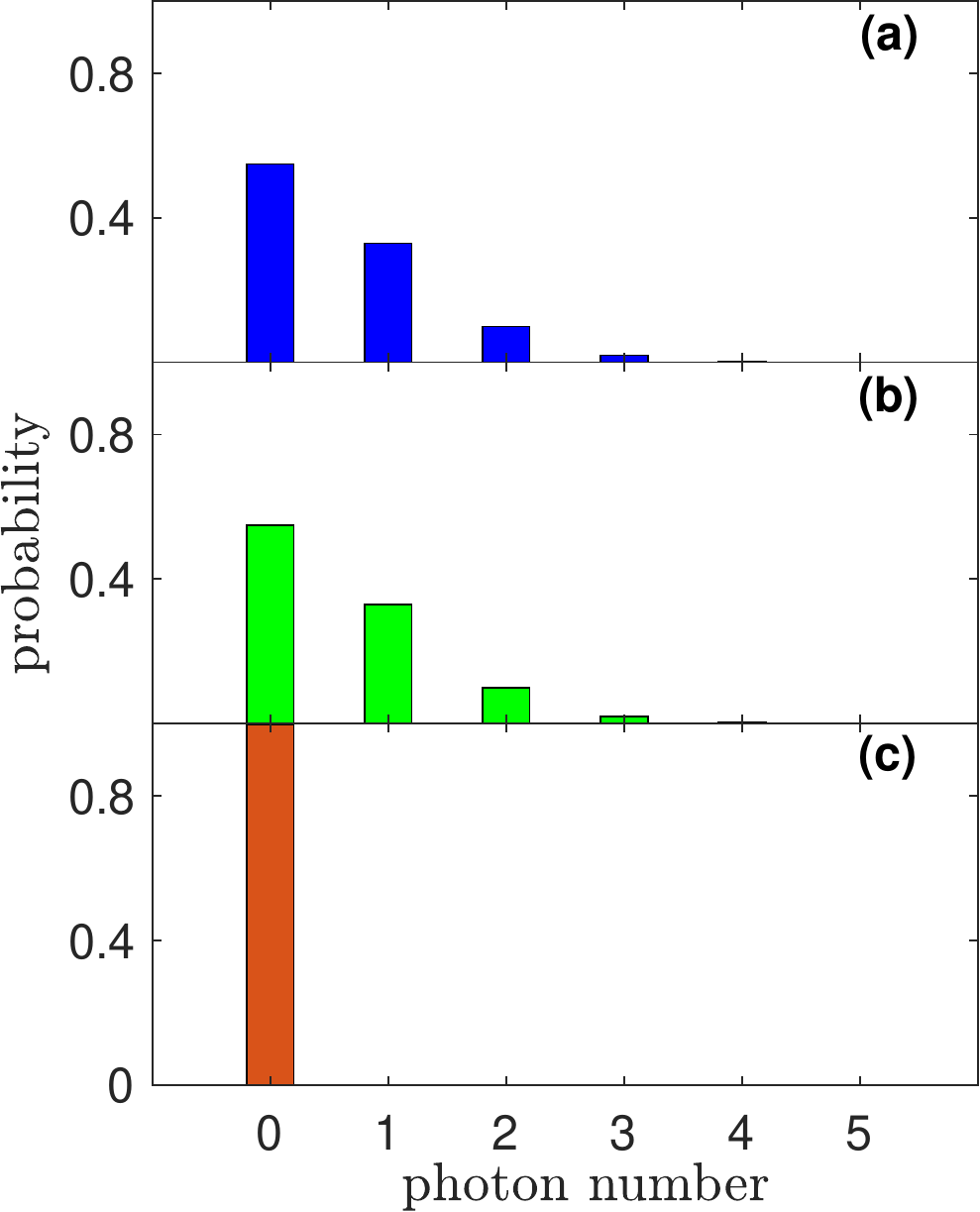}
\caption{The Fock-state probabilities of (a) the incident mode $\hat{d}$, (b) the transmitted mode $\hat{a}$ and (c) the reflected mode $\hat{c}$ of the second cavity.}
\label{fig:Fig8}
\end{figure}
In what follows we check the case that the target cavity decays only at the two ports to verify the validity of the numerical simulation of the quantum cascade method. We set $\kappa_{d1}=\kappa_{d2}$, $\kappa_{e1}=\kappa_{e2}$, $\kappa_a^\mathrm{NL}=0$. Then we extract the Fock-state probabilities of the incident mode, reflected mode and transmitted mode of the target cavity, respectively, where the reflected mode is defined as $\hat{c}=(\sqrt{\kappa_{d2}}\hat{d}+\sqrt{\kappa_{e1}}\hat{a})/\sqrt{\kappa_{d2}+\kappa_{e1}}$. The one-photon state for the incident mode is
\begin{equation}
\begin{split}
|1_d\rangle\langle1_d|&=|1_d0_a\rangle\langle1_d0_a|+|1_d1_a\rangle\langle1_d1_a|+\cdots\\
&\quad+|1_d(N_a-1)_a\rangle\langle1_d(N_a-1)_a|\\
&=|1\rangle_d{}_d\langle1|\otimes I_a,
\end{split}
\end{equation}
where $N_a$ is the truncated dimension of the Hilbert space of the target cavity, $|1\rangle_d$ denotes the one-photon state in the source cavity's Hilbert space alone and $I_a$ is the identity matrix in the target cavity's Hilbert space. Then the probability of the one-photon state of the incident mode is given by
\begin{equation}
\begin{split}
P_{1_d}&=\langle1_d|\psi\rangle\langle\psi|1_d\rangle\\
&=\text{Tr}\left\{|1_d\rangle\langle1_d|\hat{\rho}(t)\right\}\\
&=\text{Tr}\left\{|1\rangle_d{}_d\langle1|\otimes I_a\hat{\rho}(t)\right\}.
\end{split}
\end{equation}

We extract the Fock-state probabilities of the cavity mode $\hat{a}$ and reflected mode $\hat{c}$ in the same way. Fig. \ref{fig:Fig8} shows the Fock-state probabilities of the incident mode, intracavity mode (transmitted mode) and reflected mode of the target cavity. The Fock-state components of the incident mode satisfies the Poisson distribution with mean photon number $\bar{n}=0.6$, and the photon number distribution of the transmitted mode is almost the same as that of the incident mode, as shown in Figs. \ref{fig:Fig8}(a) and \ref{fig:Fig8}(b). Only the vacuum-state component in the incident field is reflected, as shown in Fig. \ref{fig:Fig8}(c). These results are consistent with the known conclusion, that is, the incident field will be totally transmitted when the decay rates of the cavity at the two ports are the same and the cavity has no intrinsic loss, which proves the validity of the numerical simulation of the quantum cascade method.

\bibliography{ref}
\end{document}